\documentclass[final,1p,times]{elsarticle}

\usepackage{graphicx}
\usepackage{amssymb}
\usepackage{amsthm}
\usepackage{lineno}

\journal{Nuclear Physics A}

\begin{document}

\begin{frontmatter}

\title{The ALICE Inner Tracking System Upgrade}

\author{Roy Lemmon (on behalf of the ALICE\fnref{col1} Collaboration)}
\fntext[col1] {A list of members of the ALICE Collaboration and acknowledgements can be found at the end of this issue.}
\address{STFC Daresbury Laboratory, Keckwick Lane, Daresbury, Warrington, Cheshire, WA4 4AD, United Kingdom}

\begin{abstract}
A central component of the ALICE Upgrade will be a completely new Inner Tracking System (ITS). The performance of the new ITS will be a significant improvement over that of the present ITS, in particular in the areas of material budget, granularity, a reduced radial distance from the first layer to the beam and rate capability. This will enable many key measurements of the properties of the quark-gluon plasma to be performed, in particular with rare probes such as low momentum charm and beauty mesons and baryons.
\end{abstract}

\end{frontmatter} 


\section{Introduction}

The ALICE Upgrade \cite{ALICE-LOI} will significantly enhance the capabilities of the experiment to study in particular rare probes of the partonic medium, the quark-gluon plasma, produced in high-energy nuclear collisions at the LHC. An overview of the ALICE Upgrade can be found in these proceedings \cite{Peitzmann12}. This paper describes in more detail the upgrade of the Inner Tracking System (ITS) \cite{ALICE-ITS-CDR}, which is a central component of the overall ALICE Upgrade. Performing new measurements of heavy-flavour production in heavy-ion collisions is one of the main physics motivations for running the ALICE experiment after the luminosity upgrade of the LHC, with extended high-rate capabilities and a new Inner Tracking System. The novel measurements of charm and beauty production would allow important questions to be addressed that cannot be answered with the present experimental setup.

The two main open questions concerning heavy-flavour interactions with the QGP medium, and the corresponding experimental handles, are:

\begin{itemize}
  \item {Thermalization and hadronization of heavy quarks in the medium, which can be studied by measuring the baryon/meson ratio for charm ($\Lambda$$_{c}$/D) and for beauty ($\Lambda$$_{b}$/B), the azimuthal anisotropy $\it{v_{2}}$ for charm mesons and baryons, and the possible in-medium thermal production of charm quarks.}
  \item {Heavy-quark in-medium energy loss and its mass dependence, which can be addressed by measuring the nuclear modification factors R$_{AA}$ of the p$_{T}$ distributions of D and B mesons separately in a wide momentum range.}
\end{itemize}

It is important to stress that the above two topics are closely related. The high-momentum heavy quarks quenched by in-medium energy loss are shifted towards low momentum and may ultimately thermalize in the system, through QCD interaction mechanisms that are essentially the same as those responsible for the energy loss, and participate in the collective expansion dynamics. Therefore, the simultaneous experimental investigation and theoretical understanding of the thermalization-related observables and of the energy-loss-related observables constitute a unique opportunity for the characterization of the QGP properties, in particular of the heavy-flavour transport coefficient. In this scope, heavy quarks are particularly well suited, because they preserve their identity (mass, flavour, and colour charge) while interacting with the medium.

\section{Detector Upgrade Concept}

The features of the ITS Upgrade as compared with the present ITS are:

\begin{itemize}
    \item {First detection layer closer to the beam line. At present the radial distance of the first layer
from the beamline is 39 mm. However, after initial studies, the installation of a new beampipe with an outer radius of 19.8mm is considered a realistic possibility. The installation of such a beampipe would enable the first detection layer to be located at a radius of about 22 mm from the main interaction vertex.}
    \item{Reduction of material budget. Reducing the material budget of the first detection layer is particularly
important for improving the impact parameter resolution. In general, reducing the overall material budget will allow the tracking performance and momentum resolution to be significantly improved. The use of Monolithic Active Pixel Sensors (MAPS) will allow the silicon material budget per layer to be reduced by a factor of 7 in comparison to the
present ITS (50 $\mu$m instead of 350 $\mu$m). A careful optimization of the analogue front-end timing specifications and readout architecture will allow the power density to be reduced by a factor of 2, despite the increased pixel density of a factor of 50. The lower power consumption and a highly optimized scheme for the distribution of the electrical power and signals will allow the material budget of the electrical power and signal cables to be
reduced by a factor of 5. Mechanics, cooling and other detector elements can also be improved when compared to the present ITS design. Combining all these new elements together, a detector with a radiation length of 0.3\% X$_{0}$ per layer or better has been demonstrated to be a realistic option. Hybrid pixels would allow the construction of detector layers with a slightly higher radiation length (0.5\% X$_{0}$), which would still represent a significant improvement of the performance as compared to the present ITS (pixel layers with 1.14\% X$_{0}$; drift layers with 1.13-1.26\% X$_{0}$; strip layers with 0.83\% X$_{0}$).}
    \item{Geometry and segmentation. The studies presented here are based on a detector consisting of seven coaxial cylindrical layers covering a radial extension from 22mm to 430mm with respect to the beamline. The physics studies of the benchmark channels are based on the assumption that all layers are segmented in pixels with an intrinsic resolution of 4 $\mu$m in both directions.}
    \item{Readout time. The present ITS features a maximum readout rate of 1 kHz. The new detector aims to read the data related to each individual interaction up to a rate of 50 kHz for Pb–Pb collisions.}
        \item{Accessibility. The mechanical design of the new ITS allows for the fast extraction and insertion of the detector for yearly maintenance. This will enable easy replacement of malfunctioning and radiation damaged modules for example.}
  \end{itemize}

A new silicon tracker featuring the characteristics listed above will enable the impact parameter resolution at the primary vertex to be improved by a factor of 3 or more (Fig. 1). The standalone tracking efficiency would be comparable to what can be presently achieved by combining the information of the ITS and the TPC. The relative momentum resolution of the upgraded ITS standalone, $\frac{\delta p}{p}$, would be about 2\% up to 2 GeV/c and remain below 3\% up to 20 GeV/c.

\begin{figure}[htbp]
\begin{center}
\includegraphics[width=0.7\textwidth]{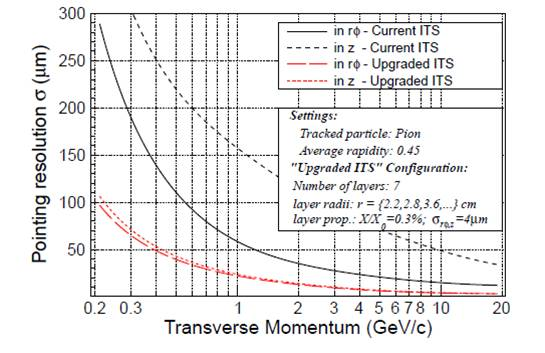}
\end{center}
\caption{Track impact parameter resolution (current and upgrade). }
\label{fig:impact}
\end{figure}

\section{Technical Implementation}

Two basic pixel detector technologies are considered for the upgrade: hybrid silicon pixel detectors and monolithic silicon pixel detectors. Investigations into the suitability of each technology are currently ongoing. For the hybrid technology the R\&D focuses on the following two key aspects: 1) the development of a new front-end chip in a CMOS 0.13 $\mu$m process aiming at reducing the power density by a factor of 2, while targeting a thickness of 50 $\mu$m; 2) the characterization of sensors with a target thickness of 100 $\mu$m, or less, and the development of low cost assembly techniques (micro bump bonding) for combining thinned frontend chips with thin sensors. The option to produce very thin sensors will be further investigated using epitaxial and thin float-zone wafers. Concerning the monolithic pixel detectors, various options are under investigation to achieve the required specifications within
the time scale of the ITS upgrade project. The pixel detectors of the MIMOSA family \cite{Greiner2010} are presently developed with a
0.35 $\mu$m CMOS process. The main limitation of these circuits is their insufficient tolerance to the radiation load expected for 10 years of ALICE operation (1.6 Mrad, 2x 10$^{13}$ 1MeV neutron equivalent). For this reason the main R\&D effort is to transfer the MIMOSA design to a smaller feature size CMOS technology (0.18 $\mu$m), which should provide a significantly higher radiation tolerance. The 0.18 $\mu$m CMOS process is a quadruple-well process that allows integrating more complex circuitry within the pixel areas. The TPAC and CHERWELL circuits are examples of such an implementation \cite{Stanitzki2010}. At present the R\&D efforts focus on the characterization of these circuits to verify their suitability to the ITS application in terms of performance and radiation tolerance.

In order to match the stringent material
budget requirements, every component of the detector module requires dedicated R\&D for its optimization. Specific emphasis will be laid on the interconnection structure of the front-end chips to the on-detector electronics which accounts for more than 40\% of the material budget in the present pixel system.
The silicon strip detectors, if upgraded, could be placed closer to the interaction vertex (e.g. first layer at a radius of 20 cm). The increased hit density (occupancy) can be handled by shortening the strip length (from currently 20 to
e.g. 10 mm). This increased readout density demands a redesign of the front-end chip and a new design of the microcables (e.g. with 50 $\mu$m pitch) interconnecting the sensor and the chip. Special emphasis will be put on low power
consumption, low material budget and a compact detector layout.

\section{Summary}

A central component of the ALICE Upgrade will be a completely new Inner Tracking System (ITS). The performance of the new ITS will be a significant improvement over that of the present ITS, in particular in the areas of material budget, granularity, a reduced radial distance from the first layer to the beam and rate capability. The new ITS will improve the impact parameter resolution by a factor of approximately 3 and allow measurements to reach much lower values of p$_{T}$. This will enable many key measurements of the properties of the quark-gluon plasma to be performed, in particular with rare probes such as low momentum charm and beauty mesons and baryons. A detailed description of the physics performance, as well as the layout options and ongoing R\&D activities can be found in the Conceptual Design Report that has recently been published. The installation and commissioning of the new ITS is planned for the long LHC shutdown period in 2017/2018. The timeplan for the upgrade foresees the R\&D and prototyping phase in 2012-2014 in order to start the production and commissioning phase in 2015.

\end{document}